# QUANTUM ASTRONOMY. PART II

By Alexander M. Ilyanok

### *Macroquantum laws in astronomy*

*A new method of fundamental quantum data sampling on basis of an optimum measuring scale has been designed. The method was applied for minimization of redundant experimental data in different fields of physics. Basing of this method there has been found the laws of binding electromagnetic, strong and gravitational interaction. It is shown that the laws of quantizing of physical quantities are the effect of space fibering. Moreover the said laws have a general electromagnetic nature as well. They are related to dimensionless electromagnetic constants - $N_\alpha = 861$ and $\alpha^{-1} = 137,0360547255$. These constants cover both the atomic and the space scales. The macroquantum laws of planets, the Sun, the Solar system, the Galaxy and the Metagalaxy as a whole have been discovered. It is demonstrated that the macroquantum laws of gravitation discretely agree with the Newton's law of gravitation. In this case the maximum velocity of gravitational interactions is equal to $\alpha^{-4}c = 3{,}526 \cdot 10^8 c$, where c is the velocity of light. It has been proved that the physics laws for condensed substance are not invariant to Lorentz-transformations, whereas the absolute traverse velocity in the Metagalaxy for condensed substance does not exceed $\alpha c = 2187{,}67$ km/s. As a matter of fact this parameter closes out the problems of interstellar traveling and space wars. A time estimation of burning of Jupiter into the second Sun shows that «the doomsday» for the mankind will come not earlier than in 50 million years. Moreover, new power sources in the core of the Earth are disclosed, which at the expense of induction essentially influence the climate of the Earth.*

## КВАНТОВАЯ АСТРОНОМИЯ
### Часть 2. Макроквантовые законы в астрономии

*Разработан новый метод фундаментальной квантовой выборки данных на основе оптимальной измерительной шкалы. Этим методом проведена минимизация избыточных экспериментальных данных в разных областях физики. На его основе найдены законы, связывающие электромагнитное, сильное и гравитационное взаимодействия. Показано, что законы квантования физических величин являются следствием расслоения пространства, а также имеют общую электромагнитную природу. Они связаны с безразмерными электромагнитными константами – $N_\alpha = 861$ и $\alpha^{-1} = 137.0360547255$. Эти константы охватывают как атомарные, так и космические масштабы. Найдены макроквантовые законы планет, Солнца, солнечной системы, Галактики и Метагалактики в целом. Показано, что макроквантовые законы гравитации дискретно совпадают с законом всемирного тяготения Ньютона. При этом максимальная скорость гравитационных взаимодействий равна $\alpha^{-4}c = 3.526 \cdot 10^8 c$, где $c$ – скорость света. Доказано, что законы физики для конденсированного вещества не инвариантны к лоренц-преобразованиям, так*

*как абсолютная предельная скорость движения в Метагалактике для конденсированного вещества не превышает α с = 2187.67км/с. Это обстоятельство фактически запрещает межзвездные путешествия и исключает космические войны в будущем. Произведена оценка времени разгорания Юпитера во второе Солнце, вследствие чего «конец света» для людей наступит не ранее чем через 50 млн. лет. Показаны новые источники энергии в ядре Земли, которые за счет индукции существенно влияют на климат Земли.*

### *Положение в современной физике*

Проблема создания новых знаний и на их основе физической картины мира обычно решается двумя способами. Первый, классический – это формирование теоретических понятий в результате систематизации индуктивного обобщения опыта. Второй – создание гипотетических моделей реальности с их последующей экспериментальной проверкой.

На основании первого способа фактически была создана классическая физика. К значимым успехам второго способа можно отнести электродинамику Максвелла, основанную на дифференциальном исчислении; теорию групп применительно к геометрии кристаллов (группы симметрии Федорова, созданные еще в прошлом веке), а также приложение группы симметрии SU(3) к спектроскопии и свойствам адронов на основе теории кварков М. Гелл-Мана, Г. Цвейга, 1964 г. [1,Т.2,с.342].

На базе второго способа были созданы: квантовая механика (КМ), специальная (СТО) и общая (ОТО) теории относительности. При этом в КМ используется Гамильтонов подход, взятый из классической механики, а в СТО и ОТО – геометрический подход.

К сожалению, настоящие теории до сих пор не смогли объяснить большинство новых экспериментальных фактов как в физике, так и в астрофизике. В основе этих теорий, как более полно будет показано, лежат ошибочные постулаты.

В общем, как отмечает Р. Фейнман в своих лекциях: «Положение, в котором находится современная физика, следует считать ужасным. Я бы подытожил его такими словами: вне ядра мы, видимо, знаем все; внутри него справедлива квантовая механика, нарушение ее принципов там не найдено». Далее он говорит: "Если когда-нибудь удастся "разгромить" принцип неопределенности, то квантовая механика начнет давать несогласованные результаты и ее придется исключить из рядов правильных теорий явлений природы".

Однако Фейнман не учитывает, что принцип неопределенности Гейзенберга фактически запрещает расширение КМ на макроквантовый уровень, то есть космос, и в этом случае КМ – неправильная теория.

В этих же лекциях Фейнман говорит: «Сцена, на которой действуют все наши знания, – это релятивистское пространство-время; не исключено, что с ним связано и тяготение». Он обратил внимание на тот факт, что «Эйнштейн взял идею Пуанкаре об инвариантности законов физики к преобразованию Лоренца». Но при этом Эйнштейн отбросил сомнения Пуанкаре, который, основываясь на глубоких знаниях проблемы в целом, перестал развивать это направление. Пуанкаре говорил: «Масса – коэффициент инерции – возрастает вместе со скоростью. Следует ли из этого заключить, что масса – коэффициент притяжения – также возрастает вместе со скоростью и остается пропорциональной коэффициенту инерции или же, что этот коэффициент притяжения остается постоянным? Это тот вопрос, решить который у нас нет никакой возможности». [2, с.507].

Более экзотические теории типа "Великого объединения" исходят из представления о единой природе сильного, слабого и электромагнитного взаимодействий, возможно, и гравитационного. Делаются попытки найти решение путем привлечения группы симметрии SU(5), а также увеличения размерности пространства от 3-х мерного до 4-х мерного и вплоть до 27-мерного. Правда, в этом случае не учитывается теорема Бертрана–Эренфеста, из которой следует, что сохранение устойчивости систем, взаимодействующих по закону $1/R^2$, возможно только при размерности пространства не более 3 [3–5].

Увлечение гипотетическим моделированием при недостаточной базе современных знаний и приписывание отдельным, часто ошибочным гипотезам выдающихся физиков статуса абсолютной истины разрушает общую методологию науки. Этим мы нарушаем одну из заповедей Библии: "Не сотвори себе кумира" – и, следовательно, переходим на субъективные метафизические позиции познания.

Все наши знания базируются на опыте. Он является единственным критерием научной истины. Законы гравитации и электромагнетизма были получены на обобщении большого числа опытов и не следовали из каких-либо философско-теоретических соображений. В общем, законы Природы, установленные нами, – это своеобразное приближение, определяемое точностью тех приборов, которые мы используем, и достоверностью тех математических моделей, которые мы взяли для описания этих законов. Как говорил Р. Фейнман в своих лекциях: «С общей точки зрения любой приближенный закон абсолютно ошибочен. Это характерное свойство общей картины мира, которая стоит за законами».

### *Введение в новый метод получения знаний*

Для описания физической реальности в виде законов необходимо создать теорию, в которой отсутствуют свободные параметры. В такой теории, в частности, должны быть получены все безразмерные константы, характеризующие физический мир. Более того, теория должна доказать единственность таких значений, например, в рамках единственности системы аксиом, опирающихся не столько на эмпирический, сколько на теоретический уровень. Или как сказал Эйнштейн: "Природа устроена так, что можно установить настолько сильные законы, что в этих законах будут только рационально полностью определенные константы (а не константы, величину которых можно менять, не разрушая теорию)" [5].

Предполагается, что в фундаментальной теории может быть выделено некоторое количество безразмерных (подлинно независимых) констант, которое никакой теорией не сможет быть уменьшено. Эти константы должны рассматриваться как фиксированные параметры физического мира.

Опираясь на вышеизложенное, поставим перед собой цель найти в виде безразмерных констант такие параметры физического мира, которые объединяют электромагнитное, гравитационное и сильное взаимодействия. При этом учтем, что электромагнитное и слабое взаимодействие уже объединены.

Найти новые безразмерные константы можно, например, путем анализа и обобщения теоретических и экспериментальных данных по модели атома Бора

[1, Т1, с.146]. По этой модели электрон, двигающийся по орбите вокруг ядра, имеет радиус:

$$r_0 = \left[\frac{n_0^2 r_e}{z\alpha}\gamma\right]_{n_0 = z = 1} = \frac{N_\alpha}{2\pi} r_e, \qquad (1)$$

где $n_0$ – главное квантовое число; $r_e = \dfrac{\hbar}{m_e c}$ – комптоновский радиус электрона, здесь $\hbar = \dfrac{h}{2\pi}$ – постоянная Планка, $c$ – скорость света; $\gamma = \sqrt{1 - \dfrac{v_0^2}{c^2}}$ – релятивистская поправка массы электрона $m_e$, здесь $v_0 = \dfrac{n_0 \hbar}{m_e r_0}\gamma = \dfrac{cz\alpha}{n_0}$ – орбитальная скорость электрона; $z$ – заряд ядра; $\alpha = \dfrac{z_0}{4\pi}\dfrac{e^2}{\hbar} = \dfrac{e^2}{4\pi\varepsilon_0 \hbar c}$ – постоянная тонкой структуры, здесь $z_0 = \sqrt{\dfrac{\mu_0}{\varepsilon_0}}$ –волновое сопротивление вакуума, $\mu_0$ и $\varepsilon_0$ – магнитная и диэлектрическая проницаемость вакуума, $e$ – заряд электрона.

Будем считать, что некая константа $N\alpha$ должна быть целым действительным числом, ближайшее значение которого находится из (1) при $n_0=1$, $z=1$:

$$N_\alpha = \frac{2\pi}{\alpha}\sqrt{1-\alpha^2} = 861. \qquad (2)$$

Отсюда вытекает и вторая безразмерная константа:

$$\alpha^{-1} = \sqrt{\left(\frac{N_\alpha}{2\pi}\right)^2 + 1} = 137.0360547255... \qquad (3)$$

Значение (3) совпадает со значением постоянной тонкой структуры [6] $\alpha^{-1}= 137.0360$ (2) до последней (седьмой) экспериментально найденной значащей цифры. Следовательно, используя (3), можно уточнить значение постоянной тонкой структуры до любой значащей цифры, определяемой только $\pi$, и косвенно входящие в нее значения мировых констант.

Назовем $N\alpha$ поперечным квантовым числом, а $\alpha$ – продольным квантовым числом. Эти числа вытекают из чисто геометрических соображений из дискретности и расслоенности пространства и справедливы от атомарных до космических размеров.

Используя идеи Бертрана–Эренфеста, введем безразмерные константы в виде значения размерности пространства: нестационарное одномерное ($n_S=1$); стационарные пространства, имеющие замкнутые движения, – двумерное ($n_S=2$) и трехмерное ($n_S=3$); нестационарное пространство с прямолинейным разомкнутым движением – четырехмерное ($n_S=4$). Безусловно, к ряду безразмерных констант следует отнести число $\pi$. Наряду с этими константами введем новые безразмерные константы, определяющие заряд ядра: $Z= 1,2,3...92$, где $Z_L=92$ – предельный заряд ядра. Это значение вытекает из $\alpha$ и $N\alpha$ и имеет естественный предел $Z_L^*<92.18$ (уран). Кроме того, из $\alpha$ и $N\alpha$ можно найти безразмерные константы: отношение масс протона и электрона, равное 1836.157198498, константу аномального магнитного момента электрона

μ=(1+$N_\alpha^{-1}$–4/3$N_\alpha^{-2}$)μ$_B$ = 1.001159641595μ$_B$, где μ$_B$ – магнетон Бора. Так как эти константы не относятся к предмету данной работы, их вывод будет приведен в последующих публикациях. По-видимому, этих безразмерных констант будет достаточно для создания теории объединения всех взаимодействий в природе.

Опираясь на идею Бертрана–Эренфеста, будем считать Вселенную трехмерным евклидовым устойчивым пространством. Оно заполнено электромагнитными и гравитационными полями, элементарными частицами и конденсированным веществом. При этом считаем, что виртуальных частиц в «физическом вакууме» не существует.

Введем гипотезу, что все взаимодействия являются электромагнитными, а пространство – дискретно и расслоено на отдельные поверхности [7]. Точное решение уравнений, совпадающих с экспериментальными данными, возможно только на этих поверхностях. Поэтому правильные решения и достоверные измерения возможны только при совпадении с определенными дискретными значениями, находящимися на соответствующих поверхностях. Промежуточные значения, как не имеющие физического смысла, отбрасываются как избыточные.

В связи с высокой степенью автоматизации экспериментов в физике накапливается большое количество избыточных данных. Выделить из них новые знания путем уменьшения избыточности можно с помощью нового способа – квантовой фундаментальной выборкой, или иначе квантовой фильтрацией опытных данных. Из-за высокого уровня «шумовой» экспериментальной информации решить проблему Великого объединения не удается уже многие годы. Это связано с тем, что решение ищется стандартными (как первым, так и вторым) способами создания новых знаний, описанных ранее.

Для сопоставления экспериментальных данных в разных областях физики введем универсальное «виртуальное» измерительное средство – безразмерную шкалу:

$$L_\alpha = b(n_1/n_2) \cdot \alpha^n + b^*(n^*_1/n^*_2) \cdot \alpha^{n^*}, \qquad (4)$$

где $n = \pm 0, 1, 2, 3\ldots$; $n_1 = 0,1,2,3,\ldots$; $n_2 = 1,2,3,\ldots$;
$b=0, 1, (2\pi)^{-1/2}, (\pi)^{-1/2}, (2\pi)^{-1}$;
$n^* = \pm 0, 1, 2, 3\ldots$; $n^*_1 = \pm 0, 1, 2, 3\ldots$; $n^*_2 = 1,2,3,\ldots$;
$b^*=0, 1, (2\pi)^{-1/2}, (\pi)^{-1/2}, (2\pi)^{-1}$.

Эта шкала легко трансформируется в шкалу с необходимой размерностью путем умножения $L_\alpha$ на элементарный заряд $e$, момент импульса $h$ (постоянную Планка), $c$ – скорость света и др.

Понятие «виртуального» измерительного средства возникает из-за невозможности однозначно экспериментально измерить параметры элементарных частиц. Поэтому, чтобы обойти неопределенность Гейзенберга для элементарных частиц, необходимо производить измерения по ряду косвенных параметров, не воздействуя на объект измерения, то есть виртуально.

Легко заметить, что постоянную Планка можно выразить через электромагнитные константы $h = \dfrac{z_0}{2}\dfrac{e^2}{\alpha}$, но для удобства написания мы ее запишем в форме $h_\alpha = \alpha^n h$ и назовем обобщенным квантом действия. Тогда постоянная Планка является частным случаем обобщенного кванта действия при $n=0$, $b=n_1=n_2=1$.

Рассмотрим другой вариант использования шкал, основанных на постоянной тонкой структуры, для нормализации гипотез Планка. В свое время Планк, введя свою постоянную, имел в распоряжении еще две мировые константы: $G$ - гравитационную постоянную и $c$ - скорость света. Комбинируя их, он получил так называемые:

планковскую массу – $\qquad m_{Pl} = \sqrt{\dfrac{\hbar c}{G}} = 2.176 \cdot 10^{-8}$ кг; $\qquad$ (5a)

планковскую длину – $\qquad l_{Pl} = \dfrac{\hbar}{m_{Pl} c} = 1.616 \cdot 10^{-35}$ м; $\qquad$ (5b)

планковское время – $\qquad t_{Pl} = \dfrac{\hbar}{m_{Pl} c^2} = 5.39 \cdot 10^{-44}$ с; $\qquad$ (5c)

планковскую энергию –

$$W_{Pl} = m_{Pl} c^2 = 1.956 \cdot 10^9 \text{ Дж } (1{,}22 \cdot 10^{28} \text{ эВ}). \qquad (5d)$$

Однако в то время не была известна постоянная тонкой структуры и не были известны значения предельных энергий в природе. В настоящее время достоверно установлено, что максимальное значение энергии космических элементарных частиц составляет порядка $10^{20}$ эВ [1, Т.2, с.471].

Для примера, не нарушая идеи Планка, путем простой нормировки $h_\alpha = 2\pi\alpha^8 \hbar$ преобразуем уравнение (5d) в соответствие с экспериментальными данными:

$$W^*_{Pl} = \sqrt{\dfrac{\alpha^8 c^5 h}{G}} = 8.68 \cdot 10^{19} \text{ эВ} \qquad (6)$$

В этом случае теория хорошо согласуется с экспериментом, что позволяет ввести соответствующие поправки в стандартную теорию Великого объединения.

Аналогичным образом можно поступить и с планковской длиной (5b), и планковской массой (5a), и с планковским временем (5c).

Следовательно, таким способом безразмерной перенормировки можно приводить теорию к эксперименту, а с помощью обобщенных шкал измерять параметры от размеров ядра, атома, конденсированного состояния вещества вплоть до масштабов Метагалактики.

В существующих курсах и учебниках по астрономии и космологии в большинстве случаев отсутствует постановка проблемы парадоксов и необходимости их решения. Понятно, что каждый парадокс – это новое заблуждение или новая физика. Без понимания этого невозможно говорить о дальнейшем развитии астрономии как науки.

Попытаемся, опираясь на наше виртуальное измерительное средство, преодолеть существующие парадоксы и получить новые знания о законах Вселенной.

***Макроквантовые законы семьи Солнца***

Рассмотрим для начала макроквантовые зависимости (законы) в солнечной системе.

Опуская промежуточные выкладки, определим основные параметры орбит планет.

Представим Солнце как плазменную сферу, состоящую из вращающихся протонов и электронов. Представим, что их электростатическое поле вращается вместе с ними. Определим предельные (относительно центра Солнца) расстояния, на которых скорость конца радиус-вектора электростатического поля, окружающего протон, будет равняться скорости света. Предположим, что минимальная скорость движения протона относительно плазмы не может быть меньше $v_p = \alpha^4 c$. Тогда предельная частота прецессии протона вокруг своей оси

$$f_p = \frac{v_p}{\lambda_p} = \alpha^4 c \frac{m_p (\alpha^4 c)^2}{hc} = 5.1732 \cdot 10^{-3} \text{ Гц}, \qquad (7)$$

где $\lambda_p = \dfrac{hc}{m_p (\alpha^4 c)^2}$ – обобщенная предельно достижимая длина волны де Бройля протона,

$m_p$ – масса протона.

Отсюда найдем предельный радиус-вектор кулоновского поля протона:

$$R_1 = \frac{c}{f_p} = \frac{h}{\alpha^{12} m_p c} = 5.795 \cdot 10^{10} \text{ м}. \qquad (8)$$

Максимальное значение большой полуоси орбиты первой от Солнца планеты – Меркурия – составляет $R_1^* = 5.791 \cdot 10^{10}$ м [6]. Различие между экспериментальными данными и расчетным значением (8) составило 0.086%.

Следовательно, значение (8) возьмем как первую реперную точку – начало шкалы измерения орбит планет.

Шкала средних орбитальных скоростей планет также начинается со значения скорости Меркурия и равняется:

$$v_1 = 3\alpha^2 c = 47.89307 \text{ км/с}. \qquad (9)$$

Это значение отличается от экспериментально измеренного значения средней орбитальной скорости Меркурия $v_1^* = 47.89$ км/с [6] всего на 0.015%, что для астрономических измерений считается абсолютным совпадением.

Теперь перейдем к электронной составляющей Солнца. Если считать, что электрон также вращается вместе со своим электростатическим полем, а за счет того, что он более легкий, чем протон, его предельная скорость не превышает $v_e = \alpha^3 c$, то предельная частота прецессии вокруг своей оси составит:

$$f_e = \frac{\alpha^3 c}{\lambda_e} = \frac{\alpha^{11} m_e c^2}{h} = 3.861 \cdot 10^{-4} \text{ Гц}, \qquad (10)$$

где $\lambda_e = \dfrac{hc}{m_e (\alpha^4 c)^2}$ – обобщенное предельно достижимое значение де Бройлевской длины волны электрона.

Тогда предельный радиус-вектор для электростатического поля электрона

$$R_5 = \frac{c}{f_e} = \frac{h}{\alpha^{11} m_e c} = 7.7647 \cdot 10^{11} \text{ м}. \qquad (11)$$

Максимальное значение большой полуоси орбиты пятой от Солнца планеты – Юпитера – составляет $R_5^* = 7.783 \cdot 10^{11}$ м [6]. Различие между экспериментальными данными и расчетным значением (11) составило 0.23%.

Увеличение погрешности для Юпитера по сравнению с Меркурием в 2.7 раза связано с увеличением в 13 раз расстояния от Солнца до Юпитера по сравнению с расстоянием от Солнца до Меркурия, что приводит к систематической ошибке измерения расстояний с Земли. Увеличение погрешности может быть также связано и с возмущениями, вызванными Сатурном.

Возьмем значение (11) за вторую реперную точку шкалы измерений расстояний в солнечной системе.

Из наших предположений вытекает, что вся солнечная система расслоена на два подпространства, генетически связанные с протоном (8) и электроном (11).

Объединим эти подпространства путем формирования макроквантовой шкалы в виде расстояний больших полуосей орбит планет:

$$R_n = \left(\frac{n + 2(2m+1)}{3}\right)^2 R_1. \tag{12}$$

Аналогично построим шкалу для средних орбитальных скоростей:

$$v_n = \frac{3v_1}{n + 2(2m+1)}, \tag{13}$$

где $n = 1,2,3,4,5,6,7,8,9$, $m = 0,0,0,0,1,2,3,4,5$.

В этом случае расстояния и скорости являются функциями двух переменных. Многократные попытки найти законы расстояний планет солнечной системы приводят к большим погрешностям, так как они сводятся к функциям с одной переменной, как, например, хорошо известный закон Титиуса–Боде [8] $R_n = R_3(0.4 + 0.3 \cdot 2^n)$, который также является функцией одной переменной, что дает погрешность 29% для Нептуна и 96% для Плутона.

Рассчитаем по приведенным формулам (12), (13) средние орбитальные скорости и значения больших полуосей орбит и сравним с экспериментом [6]. Полученные данные приведены в таблице.

| Номер планеты | Экспер. ср. орбит. скорость, $v_n^*$, км/с | Теоретич. ср. орбит. скорость, $v_n$, км/с | Погреш-ность, δ, % | Экспер. значение больших полуосей орбит, $R_n^*$, ($\times 10^6$ км) | Теоретич. Значение больших полуосей орбит, $R_n$, ($\times 10^6$ км) | Погреш-ность, δ, % |
|---|---|---|---|---|---|---|
| 1 | 47.89 | 47.893 | +0.0064 | 57.90 | 57.95 | + 0.10 |
| 2 | 35.03 | 35.919 | +2.50 | 108.20 | 103.02 | - 5.0 |
| 3 | 29.79 | 28.74 | -3.60 | 149.6 | 160.97 | +10.76 |
| 4 | 24.13 | 23.95 | -0.75 | 227.9 | 231.80 | +1.73 |
| 5 | 13.06 | 13.0617 | +0.013 | 778.3 | 779.11 | + 0.120 |
| 6 | 9.64 | 8.980 | -7.30 | 1427.0 | 1648.36 | +15.50 |
| 7 | 6.81 | 6.841 | +0.46 | 2869.6 | 2839.57 | -1.0 |
| 8 | 5.43 | 5.526 | +1.77 | 4496.6 | 4352.71 | -3.30 |
| 9 | 4.74 | 4.635 | -2.20 | 5900.0 | 6187.81 | +4.90 |

Анализ приведенной таблицы показывает, что параметры орбит планет солнечной системы строго заданы на квантовом уровне и, следовательно, генетически связаны с Солнцем, то есть они «родились» совместно с Солнцем или из самого Солнца. Отсюда следует, что между орбитами Марса и Юпитера не должно существовать крупных планет. Незначительное отклонение параметров некоторых орбит от их экспериментального значения говорит о факте обмена некоторыми из них дополнительными импульсами. Такая ситуация возможна только в случае их совместного сжатия/расширения, а следовательно, изменения температуры или посредством взаимодействия (обмена) своими спутниками.

Важно отметить, что в предельном случае уравнение квантового движения планет можно выразить и через уравнение Ньютона [1, Т2,с.474]:

$$v_n^{*2} R_n^* = GM_\Theta. \qquad (14)$$

Заменяя в этом уравнении экспериментальные значения скоростей и больших полуосей орбит на теоретические (12), (13), получим квантовое значение массы Солнца:

$$M_\Theta = \frac{9\hbar c}{\alpha^8 m_p G} \qquad (15)$$

В XIX в., когда еще не было известно о существовании протонов и электронов, Вебер и др. [2] представляли гравитацию как разницу суммарных полей положительного и отрицательного элементарных зарядов или как интерференцию положительных и отрицательных волн с нескомпенсированным остатком порядка $10^{-35}$. Развивая идеи Вебера, Хевисайд доказал, что теория тяготения должна содержать два поля и только в этом случае она подчиняется уравнениям Максвелла [9].

Следуя идеям Вебера, будем считать, что в атомах заряд протона не полностью экранируется зарядом электрона на величину $\alpha^8 e$. Тогда представим атом как сферу с нескомпенсированными электростатическими положительным и отрицательным полями.

Опуская подробности расчета электромагнитной энергии суперпозиции таких полей, с учетом только массы протонов, и пренебрегая малой массой электрона, получим следующее выражение для гравитационной постоянной:

$$G = \frac{\beta}{2\pi\varepsilon_0}\left(\frac{\alpha^8 e}{4\pi m_p}\right)^2 = 2\beta\hbar\alpha c\left(\frac{\alpha^8}{4\pi m_p}\right)^2 = 6.75334\cdot 10^{-11}\beta \text{ м}^3\text{кг}^{-1}\text{с}^{-1} \qquad (16)$$

где при поправочном множителе $\beta = 1 - \frac{4\alpha}{\pi}$ получаем окончательное значение для гравитационной постоянной $G$=6.6906·$10^{-11}$м$^3$кг$^{-1}$с$^{-2}$.

Из-за сложности экспериментального измерения $G$ существует разброс в справочных данных после третьей значащей цифры. Остановимся на данных 1982 г.: $G^*$=(6.6726±0.0005) ·$10^{-11}$ м$^3$кг$^{-1}$с$^{-2}$[1, Т.1, с.523]. Это значение расходится с теоретическим на 0.26%. Безусловно, в (16) достаточно произвольно введен поправочный множитель β, но он должен вытекать из несферичности атома и конечных пределов интегрирования нескомпенсированного электростатического поля. Однако, если в дальнейшем эксперименты покажут справедливость выражения (16), то гравитационную постоянную можно исключить, заменив ее электромагнитными константами $e$, $\varepsilon_0$, $\alpha$ и составляющей компонентой сильного взаимодействия $m_p$.

Из вышеприведенного однозначно вытекает связь электромагнитных, сильных и гравитационных взаимодействий. Эта связь также следует из соотношения параметров радиусов больших полуосей орбит Юпитера и Меркурия, которую можно получить, разделив (11) на (8), и записать в следующем виде:

$$\frac{R_5}{R_1} = \alpha \frac{m_p}{m_e} = 13.3987 . \qquad (17)$$

Значение отношения экспериментальных данных $R_5^*/R_1^*=13.442$, погрешность с теоретическим значением составляет 0.32%.

Из вышеизложенного следует очень важный вывод: электромагнитная волна, по крайней мере в солнечной системе, не может иметь длину большую, чем предельные длины волн де Бройля – электрона $\lambda_e$ и протона $\lambda_p$. Тогда частота прецессии электрона вокруг своей оси будет являться предельной минимальной частотой для солнечной системы и будет равна: $f_e = 3.861 \cdot 10^{-4}$ Гц.

Этот вывод идет вразрез с теорией Максвелла, в которой нет ограничений на максимальную и минимальную энергии электромагнитных волн. В этом смысле теория Максвелла – идеализирована, и не отвечает физическим ограничениям, реально существующим в природе, и ее следует доработать.

### *Макроквантовые законы Солнца*

Перейдем к рассмотрению макроквантовых законов самого Солнца и найдем его генетические связи с порожденными им планетами. Представим механизм рождения звезд как некий разогрев и вскипание жидкого водорода с небольшими добавками растворенного в нем гелия. В этом случае газообразный водород и гелий будут концентрироваться на внешней оболочке и в центре сферы, то есть образуется своеобразный пузырь из жидкой оболочки, заполненной газом [10].

Не вдаваясь в анализ происхождения самой материи во Вселенной и механизма первоначального толчка, стимулирующего разогрев звезды, рассмотрим модель образования только нашей ближайшей звезды – Солнца.

Введем условия неизменности объема Солнца как полого объекта:

$$V = \frac{4\pi}{3\alpha^2} R_\Theta^3 \left(1 - \frac{r_1^3}{R_\Theta^3}\right) = \text{const}, \qquad (18)$$

где $R_\Theta$, $r_1$ – соответственно внешний и внутренний радиусы оболочки Солнца. Отсюда следует:

$$\frac{r_1}{R_\Theta} = \left(1 - \alpha^2\right)^{\frac{1}{3}}. \qquad (19)$$

Из (19) находим, что толщина оболочки Солнца равна:

$$\Delta R_\Theta = \alpha^{\frac{2}{3}} R_\Theta = \frac{1}{26.5802} R_\Theta . \qquad (20)$$

При таких условиях средняя плотность солнечной оболочки составит 12.97 г/см³, что в 9.21 раз превышает расчетную среднюю плотность Солнца по всему объему и в 417 раз превышает плотность жидкого водорода в двойной критической точке, которая равна 0.0031 г/см³.

Из теории ньютоновского потенциала известно, что гравитационный потенциал при переходе через любую замкнутую оболочку претерпевает скачок в 4π, другими словами, поверхностный интеграл от вектора напряженности гравитационного поля, созданного массой $M$ по замкнутой поверхности $S$, охватывающей эту массу, равен 4π$GM$ [2].

Так как солнечное вещество по существу представляет собой низкотемпературную водородную плазму, то появляется возможность движения электронов относительно протонов. Скорость движения электронов в этом случае будет ограничена условием конденсации плазмы, то есть скоростью освобождения или второй космической скоростью для Солнца:

$$v_{\odot II} = \frac{\alpha c}{\sqrt{4\pi}} = 617.13 \text{ км/с}. \qquad (21)$$

Отметим, что экспериментально измеренная скорость освобождения на поверхности Солнца равна 617.7 км/с, что дает ошибку по отношению к расчетному значению всего в 0.09%.

Естественно, для любого конденсата, в том числе и для конденсированной плазмы, необходимо условие замкнутости "орбит" электронов. Такое условие может возникнуть на внутренней поверхности оболочки Солнца вследствие исчезновения гравитационного потенциала внутри Солнца из-за его зависимости от расстояния по закону $1/R$. В силу этого обстоятельства электроны могут двигаться по круговой орбите вдоль внутренней поверхности оболочки с первой космической скоростью:

$$v_{\odot I} = \frac{\alpha c}{\sqrt{8\pi}} = 436.381 \text{ км/с}. \qquad (22)$$

Движение электрона вдоль внутренней оболочки Солнца за счет взаимодействия с протонами приведет к кинетическому разогреву оболочки до температуры:

$$T_\odot = \frac{m_e v_{\odot I}^2}{2k} = 6282.10 K, \qquad (23)$$

где $k$ – постоянная Больцмана.

Цветовая температура в центре диска Солнца для спектра со сглаженными неоднородностями $I_\lambda$ = 6270.0K [6], то есть эта температура характерна для любой точки на Солнце. Погрешность по отношению к расчетному значению температуры составляет 0.31%. Это говорит о конечной теплопроводности оболочки Солнца.

Прямым доказательством движения электронов относительно протонов по внутренней оболочке Солнца является возбуждение ими гелиосейсмических волн. Время движения $t_1$ такой волны вдоль внутренней поверхности Солнца равно:

$$t_1 = 2\pi R_\odot \left(1 - \alpha^{\frac{2}{3}}\right) v_{\odot I}^{-1} = 160.43 \text{ мин}. \qquad (24)$$

Эта величина отличается от экспериментально зарегистрированных гелиосейсмических волн на поверхности Солнца длительностью 160.01 мин. всего на 0.26% [11].

Поперек оболочки Солнца также могут распространяться волны со скоростью $v_n = v_{\odot I}(2n+1)^{-1}$. Для второй моды $n=2$ получаем время прохождения волны:

$$t_2 = 5\alpha^{\frac{2}{3}} R_\Theta \mathrm{v}_{\Theta 1}^{-1} = 5.00 \text{ мин.} \qquad (25)$$

Здесь мы получаем так называемые пятиминутные волны, наблюдаемые на поверхности Солнца, причем расчетные и экспериментальные данные не расходятся до последней экспериментальной значащей цифры [11].

Другим прямым доказательством строения Солнца в виде тонкой сферической оболочки является двойной электрический слой электронно-протонной плазмы, отражение которой мы наблюдаем на высоте порядка 2 тыс. км в переходной области между хромосферой и короной по всей поверхности Солнца. В этой переходной области температура увеличивается по сравнению с температурой поверхности Солнца (23) до температуры $10^6$ К [6].

Попытки объяснения разогрева короны с помощью ударных, акустических, или магнитных волн, генерируемых на поверхности Солнца, не обоснованны, так как они в принципе не могут дать резкого скачка температуры на своем фронте из-за его размытости. Следовательно, резонно предположить, что этот переходной слой является результатом локального скачка двойного электрического слоя, который должен распространяться на высоту:

$$R_h = \frac{\Delta R_\Theta}{4\pi} = \frac{\alpha^{\frac{2}{3}}}{4\pi} R_\Theta = 2083.7 \text{ км.} \qquad (26)$$

Это значение дает хорошее совпадение с экспериментальными данными по высоте переходного слоя 1900…2100 км [6].

Еще одним макроквантовым законом Солнца является его период вращения $P_1$ вокруг собственной оси. Приведем его здесь без вывода:

$$P_1 = \frac{2\pi R_\Theta}{\mathrm{v}_\Theta} = \frac{16\pi R_\Theta}{\alpha^2 c} = 25.364 \text{ суток,} \qquad (27)$$

где $\mathrm{v}_\Theta = \alpha^2 c/8 = 1.995525$ км/с – экваториальная скорость вращения поверхности Солнца.

Экспериментальные данные по скорости вращения поверхности Солнца связаны со способом измерения этой скорости. Наиболее достоверными являются прямые измерения по волокнам, короне, магнитному полю или по моменту количества движения и составляют $P^* = 25.38$ суток [6], что дает погрешность по отношению к расчетному значению в 0.063% – абсолютное совпадение для астрономии.

Прекрасной иллюстрацией квантовой связи атомных молекулярных и космических величин является вращение полых сферических молекул углерода $C_{60}$. Это так называемые фуллерены, которые имеют радиус $r_F=3.17$Å. Недавно найдено, что при нагревании порошка из фуллерена выше 250К его молекулы начинают вращаться вокруг своей оси [12]. Частота их вращения при 300К составляет $f_F=10^{12}$с$^{-1}$. Отсюда можно найти «экваториальную» скорость поверхности молекул:

$$\mathrm{v}_F = 2\pi r_F f_F = 1.99177 \text{ км/с.} \qquad (28)$$

Это значение совпадает с теоретическим значением экваториальной скорости вращения Солнца $\alpha^2 c/8$ с точностью 0,19%. В этом смысле фуллерены являются своеобразной моделью кинетики звезд.

***Макроквантовые законы солнечной системы***

Естественно предположить, что следующим значением квантовой шкалы измерения скоростей волн и движений на Солнце после кванта $\alpha c$ и $\alpha^2 c$ будет квант $\alpha^3 c$. Следовательно, этой скорости можно задать и макроквант первой космической скорости $v_2$. По аналогии с (22) запишем:

$$v_2 = \frac{\alpha^3 c}{\sqrt{8\pi}}. \qquad (29)$$

Предположим, что каждой планете, двигающейся вокруг Солнца, будет соответствовать волна, двигающаяся синхронно в оболочке Солнца. Естественно, что эта волна, аналогично лунным приливным волнам на Земле, имеет два выступа, симметрично расположенных по оси, соединяющей Солнце и планету. Если такая волна (двойной солитон) двигается по внутренней оболочке Солнца, то ее удвоенный период $P_2$ составит:

$$P_2 = \frac{4\pi R_\Theta \left(1 - \alpha^{\frac{2}{3}}\right)}{v_2} = 11.456 \text{ лет}. \qquad (30)$$

Этому периоду соответствует среднее значение солнечной активности, равное 11.04 г., которое лежит в диапазоне от 7.5 до 16 лет. Это значение также близко к периоду обращения Юпитера вокруг Солнца, равному 11.86223 года. Расхождение в 3.5% говорит о том, что, по-видимому, солитонные волны двигаются ближе к внешней поверхности оболочки Солнца.

Рассматривая все остальные планеты, можно предположить, что каждой из них соответствует волна на Солнце с периодом:

$$P_n = \frac{2\pi R_n}{v_n} = \frac{2\pi R_1}{v_1} \left[\frac{n + 2(2m+1)}{3}\right]^3 \qquad (31)$$

Как следует из (12), орбиты планет существенно не изменяли своих параметров за миллиарды лет своего движения, так как мировые константы не могут изменяться. Следовательно, энергия приливных волн на Солнце, вызванных планетами, не поглощается. Поэтому существует обратное явление – передача энергии приливных волн от Солнца планетам. Такой механизм определяет устойчивость солнечной системы до момента полного "выгорания" энергии Солнца и падения планет на него.

Покажем, что наличие полости в Солнце следует также из законов сохранения кинетической энергии вращения Солнца и кинетической энергии поступательного движения планет.

Парадокс несоблюдения законов полного момента количества движения и полной кинетической энергии солнечной системы обсуждается на протяжении уже не одного столетия.

Со времен Ньютона вопрос о механизме начального толчка движения планет оставался открытым. Только в 1960 г. Эдьед нашел решение на основе старой гипотезы Дирака об изменении гравитационной постоянной со временем. Он установил, что первоначально Солнце являлось звездой-гигантом, которая при сжатии периодически сбрасывала свою массу в виде планет, начиная с Плутона, передавая им начальный импульс движения [13].

Аналогично теории Эдьеда предположим, что Солнце в начальный момент было звездой-гигантом, при этом предположим существование внутри него полости. Также предположим, что полое Солнце под действием гравитационных сил находится в состоянии упругого сжатия. Тогда такую систему можно описать дифференциальным уравнением второго порядка по аналогии с упругими полыми сферами, рассматриваемыми в классической механике и находящимися под внешним равномерным давлением [11]. Если представить, что роль внешнего равномерного давления выполняет гравитационное взаимодействие между частицами оболочки Солнца, то дифференциальное уравнение имеет решения в виде: равномерного движения тела, его вращения вокруг оси и движения волн в оболочке. Все эти движения присущи Солнцу. В последнем случае потенциальная энергия упругого гравитационного сжатия переходит в кинетическую энергию незатухающего движения волн по оболочке, которая, в свою очередь, посредством гравитационных взаимодействий передается планетам. В этом случае любое возмущение на внутренней экваториальной поверхности Солнца может породить солитон – каплю материи, которая будет являться зародышем будущей планеты. При сжатии оболочки Солнца этот зародыш сохраняет импульс движения и остается на заданной орбите. В этом случае гравитационный потенциал планеты по отношению к внутренней поверхности оболочки Солнца претерпевает скачок в 4π, аналогично скачку потенциала при переходе через двойной электростатический или двойной гравитационный слой.

Кинетическая энергия Солнца, представленного в виде вращающейся сферы с массой, сосредоточенной в основном в оболочке, находится следующим общепринятым способом для полой сферы:

$$W_k = (M_\Theta \, v_\Theta^2)/3 = 2.64 \cdot 10^{36} \text{ Дж}, \qquad (32)$$

где $v_\Theta$ – экваториальная скорость вращения поверхности Солнца.

Кинетическая энергия движения всех планет солнечной системы без учета кинетических энергий вращения планет вокруг собственных осей и энергии движения Плутона равняется [6]:

$$W_p = \frac{1}{2} \sum_{n=1}^{8} M_n v_n^2 = 1.99 \cdot 10^{35} \text{ Дж}, \qquad (33)$$

где $M_n$ – масса $n$-ой планеты; $v_n$ – орбитальная скорость $n$-ой планеты.

Учтем, что при рождении планеты и переходе через солнечную оболочку ее зародыша происходит скачок гравитационного потенциала в 4π. Приравнивая (32) и (33) с учетом этого скачка, получим для закона сохранения кинетической энергии в солнечной системе следующее выражение [11]:

$$\frac{M_\Theta v_\Theta^2}{3} = 2\pi \sum_{n=1}^{8} M_n v_n^2 . \qquad (34)$$

Погрешность при расчете по этой формуле составляет 5.4%.

Закон сохранения энергии солнечной системы (34) показывает, что наша солнечная система не могла сталкиваться с другими звездами после момента ее рождения и не имеет других достаточно крупных планет, кроме известных восьми. Кинетическая энергия вращения планет, как следует из выражения (34), не связана генетически с процессом рождения планеты из Солнца, а, по-видимому, связана с энергией гравитационного сжатия планет.

Закон сохранения полного момента количества движения в солнечной системе без доказательства следует из следующего выражения:

$$\frac{2}{3}M_\Theta R_\Theta v_\Theta = \frac{1}{2n+1}\sum_{n=1}^{8} M_n R_n v_n = \frac{\alpha^2 c R_1}{17}\sum_{n,m} M_n[n+2(2m+1)]. \quad (35)$$

Расчет по этой формуле дает ошибку в 1.6%.

Рассмотрим, как квантуются сами массы в солнечной системе. Экспериментальные значения отношения массы Солнца к суммарной массе всех планет и их спутников за исключением Юпитера, а также астероидов и комет составит [6]:

$$N^*_\Theta = \frac{M_\Theta}{\sum_i M_i} = 2580.08. \quad (36a)$$

Если использовать шкалу на основе $N_\alpha$, то ближайшее теоретическое значение можно найти по формуле:

$$N_\Theta = 3(N_\alpha - 1) = 2580.00, \quad (36b)$$

что совпадает с экспериментом.

Большое количество квантовых эффектов в солнечной системе было найдено Молчановым [14]. Он назвал их резонансами, которые подчиняются уравнению:

$$\sum_i n_i \omega_i = 0, \quad (37)$$

где $n_i$ – целые положительные числа и нули; $\omega_i$ – средняя частота вращения планет. В этом случае резонансы являются естественным квантовым состоянием солнечной системы. Их также можно связать с квантом действия по предложенной выше схеме.

Рассмотрим, как на основе предлагаемой методики можно устранить некоторые противоречия основных теорий гравитации.

Остановимся на сравнении двух теорий гравитации. Первая – теория гравитации Ньютона, в которой гравитационные взаимодействия распространяются с бесконечно большой скоростью, что приводит к парадоксу – нарушению закона сохранения энергии поля. Вторая – теория Эйнштейна – ОТО, в ней скорость распространения гравитационных взаимодействий ограничена скоростью света. Здесь также нарушаются законы сохранения энергии.

Предполагается, что, по сравнению с теорией Ньютона, ОТО обладает тремя основными преимуществами: описывает гравитационное красное смещение фотонов, гравитационное отклонение фотонов и аномальное смещение перигелия Меркурия. Однако гравитационное красное смещение следует также и из теории Ньютона и было найдено еще Лапласом в XVIII в. [15]. Таким образом, в ОТО и в теории Ньютона решения по красному смещению дают одинаковые результаты.

Как было показано ранее, в частном случае макроквантовые законы сводятся к классическим ньютоновским законам (14) и (15). Следовательно, все выводы ньютоновской теории гравитации по красному смещению фотонов применимы и в нашем случае.

Гравитационное отклонение фотонов было найдено Зольднером в 1804 г. [16]. Он ответил на третий вопрос Ньютона, поставленный последним в работе "Оптика", – о гравитационном взаимодействии фотонов как неких тяжелых объектов с космическими телами. Найденная Зольднером величина отклонения фотонов в поле Солнца была в 2 раза меньше, чем в ОТО [1, Т. 5, с.188].

Как уже отмечалось в модели полого Солнца, масса сосредоточена в оболочке. Поэтому пролетающий фотон взаимодействует не с точечной массой, сосредоточенной в центре Солнца, а с массой, распределенной по оболочке в виде слоя. Предположив, что скорость распространения гравитационного взаимодействия фотонов с Солнцем конечна и равна скорости света, можно легко получить тот же удвоенный коэффициент отклонения фотонов, что и в ОТО. Из экспериментов по отклонению фотонов и радиоволн Солнцем следует, что скорость распространения гравитационного взаимодействия с электромагнитными волнами, в том числе и с фотонами, равняется скорости света. В этом случае Эйнштейн абсолютно прав.

Для других элементарных частиц и конденсированного вещества прямых экспериментальных данных по скорости взаимодействия нет. Однако Лаплас оценивал эту скорость по движению планет солнечной системы: порядка $7 \cdot 10^6$ скоростей света [2, 17].

Последним расхождением ОТО с ньютоновской теорией является тот факт, что в ОТО объясняется аномальное значение перигелия Меркурия, а в ньютоновской – нет. Если решать задачу в рамках ньютоновского потенциала, но при этом учесть, что Солнце состоит из плазмы в виде электронов и протонов, то ускорение свободного падения на Солнце можно записать следующим образом:

$$g = \frac{GM_\Theta}{R^2}\left[1 + \frac{m_e}{m_p} \cdot \frac{R_\Theta^2}{R_1^2}\right], \qquad (38)$$

где $R_\Theta$ и $R_1$ – соответственно радиусы Солнца и орбиты Меркурия.

Интегрирование этого уравнения стандартными для небесной механики способами [18] дает для Меркурия сдвиг перигелия без учета эксцентриситета орбиты Δφ = 42"32 за столетие, а с учетом эксцентриситета – 44"19. Экспериментальные данные лежат в пределах 41"4 ≤ Δφ ≤ 43"1, а теоретические расчеты в ОТО дают Δφ = 43"03 [19]. Как известно, если есть одно решение задачи, то их может быть бесконечное множество [19].

К сожалению, ни теория Ньютона, ни теория Эйнштейна не являются законченными теориями гравитации и нуждаются в разработке некоторых принципиальных вопросов. Важнейшим из них является вопрос о природе поля гравитации, в частности проблема энергии и импульса поля, гравитационных волн и гравитационного излучения. Эти фундаментальные вопросы не могут быть убедительно разрешены в рамках этих теорий, ибо они построены чисто феноменологически. Можно думать, что с указанными вопросами связана и проблема квантования гравитации, поскольку попытки построить квантовую теорию тяготения по аналогии с материальными полями оказались неудачными.

Попытка Эйнштейна устранить противоречия ньютоновской теории, ограничив скорости взаимодействия посредством гравитационных сил скоростью света, оказалась непродуктивной. Его подход можно использовать только для электромагнитного излучения, распространяющегося относительно гравитирующего объекта со скоростью света. Для элементарных частиц, молекул и конденсированных веществ прямых экспериментов по измерению скорости гравитационного взаимодействия нет [20].

В нашем случае можно представить гравитационное взаимодействие как некое электромагнитное поле (из нескомпенсированных положительных и отрицательных полей зарядов), скорость возникновения и распространения которого в пространстве равняется скорости света. После установления в пространстве такого поля скорость распространения гравитационных взаимодействий между телами будет различной. В предельном случае из соображений симметрии по отношению к (7), (10) при гравитационном взаимодействии холодных конденсируемых космических тел эта скорость не должна превысить $\alpha^{-4}c$. Чем холоднее тело и чем медленнее оно двигается в пространстве, тем с большей скоростью оно взаимодействует с другими телами. Образно говоря, между гравитирующими объектами со скоростью света натягивается некая струна или поверхность, по которой передаются волны со скоростью, значительно более высокой, чем сама скорость света. Это хорошо согласуется с оценками Лапласа $> 7{\cdot}10^6 c$ [17]. Кроме того, прямым экспериментальным доказательством превышения скорости света при взаимодействии частиц между собой являются эффекты взаимодействия между собой генетически связанных фотонов [21;1,Т.1,с.184] и резонансное поглощение ядрами γ-квантов – эффект Мессбауэра [1, Т.1, с. 100].

### *Квантовые законы Земли и некоторых других планет*

Перейдем к более близким к нам проблемам – квантовой структуре Земли. Если представить, что Земля, аналогично всем планетам, «родилась» из Солнца, то в этом случае она является кусочком звезды. Известно [22], что Земля состоит из внутреннего ядра, внешнего ядра, мантии и коры. Внутреннее ядро имеет радиус $R_{\oplus 1}$ = 1217.1 км [23]. Оно, по-видимому, состоит из газообразной субстанции, не пропускающей поперечных сейсмических (акустических) волн. Давление газа в нем составляет $3.6324 \cdot 10^2$ ГПа. Внешнее ядро имеет радиус $R_{\oplus 2}$ = 3485.7 км. Оно состоит из жидкой субстанции и имеет оценочную температуру 6200К, а плотность его составляет 13.012 г/см$^3$ [6]. Такая плотность практически полностью совпадает с плотностью солнечной оболочки, рассчитанной по формулам (18)…(20). Далее располагается полужидкая мантия радиусом $R_{\oplus 3}$ = 6031 км, покрытая тонкой твердой корой с $\Delta R_{\oplus 4}$ = 340 км.

Будем считать, что так же, как и для Солнца, внешнее ядро Земли является водородной плазмой. По его внутренней поверхности могут распространяться волны, имеющие скорость:

$$v_{\oplus 1} = \frac{\alpha\, c}{\sqrt{8\pi}}. \qquad (39)$$

Эта волна имеет период обращения вокруг ядра:

$$P_{\oplus 1} = \frac{2\pi\, R_{\oplus 1}}{v_{\oplus 1}} = 17.523 \text{ с}. \qquad (40)$$

Волна (40) в виде пика прекрасно регистрируется всеми сейсмическими станциями в океанах в диапазоне Pc3 (10–45с), а на суше кроме нее видна и вторая гармоника с периодом 8.76с в диапазоне Pc2 (2–10с) [24]. Эти пики не имели ранее никакого теоретического объяснения, хотя суммарная энергия волн в этих диапазонах превышает суммарную энергию всех сейсмических волн Земли.

Важно, что на частотах близлежащего диапазона Pc3 колеблется электромагнитное поле Земли [25], а плазмосферный резонанс приходится на область Pc3, находящуюся ниже $2 \cdot 10^3$ км. При этом волны Pc2, Pc3 являются стоячими. Существенно, что фазовая скорость геомагнитных пульсаций на поверхности Земли с периодами около 10 сек. близка к 450км/с и полностью совпадает с (39).

Здесь четко просматривается плазменная область $2 \cdot 10^3$ км, аналогичная переходному слою на Солнце (26). Следовательно, энергия движения волны во внешнем ядре Земли должна передаваться как земной коре, так и ионосфере и непосредственно влиять на климат Земли. Такое влияние можно проследить на примере океанов.

Известно [26], что в воде выведенные из равновесия частицы будут совершать малые колебания около положения равновесия на частоте Вяйсяля $f_v$. Причем частота колебаний будет зависеть как от глубины, так и от солености воды. Так, в северной части Тихого океана максимум $f_v$ находится на глубине от 100 до 200м и равен 0.02 Гц, для экваториальной области на той же глубине $f_v$ = 0.06Гц, что уже совпадает с частотой сейсмических волн Pc3 0.057Гц и частотой геомагнитных пульсаций. Следовательно, энергия волн ядра Земли резонансным способом может передаваться непосредственно верхним слоям океана в основном в экваториальной области. Таким образом, над всей поверхностью океанов образуется своеобразный тепловой щит на глубине от 100 до 200м. Этот щит спасает нас от глубинных холодных вод, имеющих среднюю температуру 3,8°С. Солнечные лучи нагревают только несколько десятков метров на поверхности воды и на глубину 100 – 200 метров они не могут проникнуть, а именно здесь сосредоточены основные источники тепла океана.

По-видимому, движение некоторых волн во внешнем ядре Земли происходит в разнообразных направлениях. Это, в свою очередь, может привести к индуцированному течению океанских вод – например в виде течения Гольфстрим.

Если обратиться к далекому прошлому Земли, то можно обнаружить, что около 110 млн. лет назад средняя температура воды океанов составляла 23°С, а ее уровень был на 500 м выше нынешнего [27]. Это нельзя объяснить падением крупных космических тел на Землю, так как пылевые облака от взрывов рассеиваются в течение считанных лет. Следовательно, вероятным источником изменения климата на Земле могут быть долгопериодические процессы в самом ее ядре или же эти изменения имеют космическое происхождение.

Известно, что светимость Солнца не менялась последние 3 млрд. лет и не могла влиять на изменение климата на Земле [27]. По-видимому, внешним космическим источником, способным повлиять на ее климат, является Юпитер. В начале XX в. некоторые астрономы считали Юпитер и все крупные планеты потухшими звездами. Но потом эту идею забыли. Однако она может вполне объяснить изменение климата на Земле, если предположить, что Юпитер способен разгораться и превращаться в действующее солнце с периодом порядка 100 млн. лет. Прямым доказательством этой гипотезы является то, что Юпитер состоит из водорода, а скорость его экваториального вращения ровно в $2\pi$ раз выше скорости экваториального вращения Солнца. Кроме того, все крупные спутники Юпитера, в отличие от спутников других планет, имеют полностью оплавленный вид. Впрочем, во всем этом предстоит еще более подробно разобраться.

Допустим, что движение электронов вдоль внутренней поверхности внешнего ядра Земли также вызывает движение их электростатического поля. Тогда концы векторов этого поля не могут двигаться в конденсированном теле (мантии) со скоростью бол́ьшей $\alpha c$. Следовательно, конец радиус-вектора такого поля будет находиться на расстоянии

$$R_{\oplus}^{'} = \sqrt{8\pi} R_{\oplus 1} = 6101.64 \text{ км.} \qquad (41)$$

То есть, радиус-вектор заканчивается на глубине 270 км от поверхности Земли.

Применяя такую модель внутреннего строения Земли, можно найти решение парадокса, связанного с механизмом разогрева мантии Земли. Сначала считалось, что ее разогрев происходит путем простой диффузии тепла из ядра Земли, которое сохранилось с момента ее возникновения. Однако из-за малости коэффициента диффузии тепла в мантии от такой модели пришлось отказаться. Тогда перешли к конвективной модели: разогрев мантии происходит за счет передачи тепла путем локального перемешивания. В этом случае конвективные потоки должны были бы вызывать постоянные сильные землетрясения, что также не наблюдается. Прибегли к модели мантии, которая разогревается за счет радиоактивного распада долгоживущих изотопов $U^{238}$, $U^{235}$, $Th^{232}$, $K^{40}$. При такой модели разогрев мантии и ядра Земли должен был дать температуру в три раза меньше наблюдаемой. Такой разогрев должен был бы привести к радиоактивному загрязнению Земли за счет вулканической деятельности, что также не наблюдается. [22]

В нашей модели разогрев происходит путем электромагнитной индукции, вызывающей тепловой разогрев мантии одновременно во всей ее толще. Это исключает механизмы диффузии и конвекции и одновременно радиоактивного загрязнения поверхности Земли.

Кроме того, небольшая диффузия водорода из внешнего ядра Земли через мантию позволяет объяснить механизм образования воды на Земле. Модель водородного ядра у Земли уже применялась ранее при попытке объяснения происхождения воды на Земле, но водородное ядро не учитывалось как источник энергии [28, с.90].

Перенесем нашу модель строения Земли на Венеру.

Возникновение планет солнечной системы будем считать идентичным. Учтем, что Венера по размерам близка к Земле, ее радиус равен $0.949 R_{\oplus}$, где $R_{\oplus}$ – радиус Земли. В этом случае конец радиуса-вектора электростатического поля заканчивается на высоте 55 км над поверхностью Венеры. Сейсмические исследования, проведенные на поверхности Венеры космическими аппаратами дают глубину коры порядка 30 км.

Отсюда следует неутешительный вывод – в будущем, в течение 1 – 3 млрд. лет, из-за высокой температуры поверхности Венеры освоение ее людьми принципиально невозможно.

Проводя аналогии с Солнцем, считаем, что за счет гравитационного сжатия внешнего ядра Земли происходит преобразование потенциальной энергии сжатия в кинетическую энергию вращения. При этом скорость вращения поверхности на экваторе Земли равняется

$$\text{v}_{\oplus} = 4\alpha^{3} c = 465.981 \text{ м/с.} \qquad (42)$$

Экспериментальное значение экваториальной скорости $v_{\oplus}^{*}$=465.10м/с, что дает по сравнению с теоретическим значением погрешность в 0.19%.

Найдем радиус Земли в момент ее «рождения». Будем считать, что первая космическая скорость на поверхности Земли квантована и равна:

$$v_{\oplus 1} = \alpha^2 c / 2 = 7.9821 \text{ км/с}. \qquad (43)$$

Этой скорости соответствует экспериментально измеренная продольная скорость распространения сейсмических волн в коре на глубине 100 км при температуре 1200К [6].

Экспериментально измеренная первая космическая скорость на поверхности Земли составляет $v_{\oplus 1}^{*}$ = 7.9125 км/с. Отсюда следует, что в момент «рождения» Земли ее радиус, соответствующий скорости (43), был меньше нынешнего на 115 км и имел величину

$$R_{\oplus} = \frac{4GM_{\oplus}}{\left(\alpha^2 c\right)^2} = 6256.07 \text{ км}. \qquad (44)$$

То есть в момент «рождения» Земли (4.5 млрд. лет назад) конец радиус-вектора электростатического поля находился на 115 км ниже поверхности, и условия на Земле были похожи на условия, существующие ныне на Венере.

Используя предложенную модель, объясним образование магнитного дипольного поля Земли.

Представим, что по внутренней поверхности внешнего ядра двигается в одном направлении кольцевая волна электронов со скоростью $v_{\oplus 1}$ (39). В противоположном направлении двигается аналогичная, но не равная по энергии волна. Оси этих двух кольцевых токов не параллельны. В этом случае суммарное магнитное поле этих токов может быть смещено относительно центра Земли. В данный момент по экспериментальным данным такое смещение составляет 462 км. Северный полюс оси диполя и Южный полюс оси диполя будут также смещены относительно оси Земли. Известно, что магнитные полюса Земли медленно дрейфуют вплоть до изменения полярности. По-видимому, это связано с движением кольцевых токов по внутренней поверхности внешнего ядра.

Такой подход, в принципе, объясняет все парадоксы, связанные с магнитным полем Земли, и указывает реальный источник энергии, его создающий.

Напомним, что кинетическая анизотропная энергия гравитационного сжатия внешнего ядра Земли переходит в энергию движения токов в плазме, в температуру Земли, кинетическую энергию вращения Земли и т.д.

Покажем, что плазменные волны внутри Земли передают энергию океанам, проявляющуюся в виде приливов и отливов.

Со времен Ньютона считалось, что приливы и отливы на Земле создаются Луной: кинетическая энергия движения Луны преобразуется в кинетическую энергию движения воды на Земле. Следуя этому предположению, Луна приближается к Земле [29]. Кроме этого, как указывал еще Лаплас, на Земле должно существовать два волновых водяных горба. Однако, как оказалось, эти горбы смещены относительно оси Земля–Луна на $2.16^{\circ}$ вперед.

Отсюда следует, что не кинетическая энергия Луны, а кинетическая энергия Земли передается Луне и, следовательно, Луна должна удаляться. Однако многократные сверхточные эксперименты по измерению расстояния до Луны не обнаружили ни удаления, ни приближения Луны. Кроме того, если подробно рассматривать фазовые зависимости приливов в мировом океане, то они больше напоминают волны в резонаторах, то есть теоретические лапласовские приливы, очевидно, не могут служить даже в качестве первого приближения при описании явлений приливов и отливов [29].

Разрешить этот парадокс можно, предположив, что Луна относительно Земли находится на строго заданном квантовом расстоянии, то есть аналогично рассмотренному нами выше квантованному расположению планет относительно Солнца и механизму их движения.

Представим, что вдоль внутренней поверхности внешнего ядра наряду с быстрыми модами (39), **(42)** существует медленная волновая мода, имеющая скорость движения:

$$v_{\oplus 3} = 4\alpha^4 c. \qquad (45)$$

Период распространения этой волны равен:
$$P_{\oplus 3} = \frac{2\pi R_{\oplus 1}}{v_{\oplus 3}} = 26.103 \text{ суток}. \qquad (46)$$

Этот период меньше реального сидерического периода (27.32166 суток) движения Луны вокруг Земли на 4.67%. Совпадение периодов можно получить, если предположить, что сама волна должна распространяться несколько глубже, а не на поверхности внешнего ядра Земли. Волна должна быть наклонена к плоскости экватора Земли в соответствии с орбитой Луны.

Отсюда можно сделать вывод, что плазменная волна (45) индуцирует на поверхности Земли движение водных масс в океанах в виде двух горбов (приливов и отливов), опережающих движение Луны.

Таким образом, отсюда следует, что гравитационная энергия сжатия внешнего ядра Земли также передается и нашему спутнику Луне. Ей предстоит быть нашей вечной спутницей вплоть до полного остывания Земли, лишь тогда она сольется с ней.

Представим без выводов еще несколько важных формул, описывающих некоторые параметры планет семьи Солнца.

Радиус Юпитера запишется в виде:
$$r_5 = \frac{GM_5 N_\alpha}{4} \cdot \left(\frac{\alpha c}{\sqrt{4\pi}}\right)^{-2} = 7.16326 \cdot 10^4 \text{ км}. \qquad (47)$$

Экспериментальные данные дают размер экваториального радиуса $r_5^* = 7.1300 \cdot 10^4$ км, погрешность составляет 0.38% [6].

Экваториальная скорость поверхности Юпитера
$$v_5 = 2\pi \frac{\alpha^2 c}{8} = 12.5383 \text{ км/с}, \qquad (48)$$

что составляет ошибку по сравнению с экспериментальной в 0.061%. Эта скорость больше экваториальной на поверхности Солнца ровно в $2\pi$ раз.

Справедливо также соотношение:

$$\frac{M_\Theta}{M_5} \cdot \frac{R_5}{R_\Theta} = \frac{N_\alpha}{8}.  \qquad (49)$$

Необходимо обратить внимание и на некоторые особенности внутренних планет солнечной системы [6]. Например, экваториальный радиус Марса, равный 3395 км, отличается всего на 2.7% от радиуса внешнего ядра Земли, равного 3485.7 км. Экваториальный радиус Меркурия равен 2425 км, что в 2 раза больше радиуса внутреннего ядра Земли (погрешность составляет 0.37%).

Это еще раз свидетельствует о совместном процессе «рождения» солнечной системы.

*Макроквантовые законы в Метагалактике*

Исторически неоднократно предпринимались попытки найти законы квантования для Вселенной, используя различные подходы. Одной из них были исследования Дирака по большим числам [30]. Он попытался найти связь между постоянной Хаббла и мировыми константами, но безуспешно. Попытки других исследователей также оказывались неудачными. Это связано с преобладающей на данное время физической картиной мира – релятивистским пространством-временем. В ней предполагается инвариантность всех законов физики относительно лоренц-преобразований. При выдвижении тезиса абсолютности скорости света и изменения масштабов расстояний масс и времени в соответствии с лоренц-преобразованиями не учитывался очень важный момент: все объемные тела практически сводились к математической точке, расположенной в центре масс тела, а уже относительно нее делались формально правильные преобразования Лоренца. При этом не учитывалось, что любое тело находится в определенном фазовом состоянии: твердом, жидком, газообразном (молекулярном, атомарном), плазменном в виде смеси элементарных частиц и ионов. Кроме того, не учитывались спиновые состояния при движении частиц в пространстве. Эти казалось бы второстепенные факторы принципиальным образом меняют интерпретацию экспериментальных данных.

Известно, что все виды наблюдений в оптическом, радио, рентгеновском и других диапазонах космических объектов выявили, что максимальная скорость движения тел и их компонентов относительно друг друга не превышает 5500 км/с (расширение оболочки сверхновой звезды Кассиопея А относительно ее центра или удвоенное значение $1.1 \cdot 10^4$ км/с для противоположных составляющих оболочки [23, с.1220]). Аналогичные скорости расширения наблюдаются и в распадающихся квазарах. Например, ширина эмиссионных линий квазара 3С273 определяется скоростью $10^4$ км/с. Для сейфертовских галактик первого типа дисперсия лучевых скоростей лежит в диапазоне $(1–3)10^3$ км/с.

Постулируем, что относительные скорости движения в космосе конденсированных сред не превышают для кругового движения

$$V_1 = 3\alpha\, c = 6563.07 \text{ км/с}, \qquad (50)$$

а для прямолинейного движения –

$$V_2 = \alpha c = 2187.69 \text{ км/с}. \qquad (51)$$

По существу, при этих скоростях двигается уже не конденсированная материя звезды, а ее молекулярные, атомарные и плазменные компоненты или гравитационно не связанные объекты. Например, лучевая скорость сейфертовской галактики NGC3227 =1200 км/с [6].

Здесь необходимо особо подчеркнуть, что галактики размером более 1 кпс становятся неустойчивыми из-за изменения закона гравитационного взаимодействия. При таких расстояниях силы гравитационного взаимодействия убывают медленнее, чем $1/R^2$. Здесь закон всемирного тяготения Ньютона перестает действовать и необходимо переходить к макроквантовым законам, учитывающим изменение $G$.

В земных условиях проводились неоднократные эксперименты по достижению больших скоростей конденсированными телами-макрочастицами. На протяжении 50–70-х гг. были неоднократные попытки получить термоядерный синтез путем разгона заряженных макрочастиц дейтерия и трития в ускорителе с последующим их торможением на мишени [31]. Однако ни в одном эксперименте не удалось превысить скорость $\alpha c/\sqrt{4\pi}$ =617.13 км/с, так как все частицы испарялись.

Почему это происходит? Известно, что при ускорении электронов их спин поворачивается по направлению движения или против [32]. При разгоне конденсированного объекта спины электронов стремятся повернуться по или против направления движения. При таком повороте спина электрона в конденсированном теле этому телу передается дополнительная энергия за счет релятивистской поправки:

$$\Delta W = m_e c^2 \left( \frac{1}{\sqrt{1-\alpha^2/4\pi}} - 1 \right) = 1.1 \text{ эВ}. \qquad (52)$$

Эта энергия характерна для плавления твердых тел. При дальнейшем увеличении скорости движения энергии частиц достигнут величины

$$\Delta W = m_e c^2 \left( \frac{1}{\sqrt{1-\alpha^2}} - 1 \right) = 13.6 \text{ эВ}. \qquad (53)$$

Полученная энергия равна энергии ионизации водорода, то есть энергии образования плазмы. Тем более, что большинство конденсированных веществ испаряется при значительно меньших удельных энергиях.

Отсюда постулируем:

*предельная скорость молекул при испарении любого конденсированного вещества равна*

$$V_3 = \frac{\alpha c}{\sqrt{4\pi}} = 617.13 \text{ км/с}. \qquad (54)$$

Этой скорости соответствуют предельные скорости движения атомов и молекул на поверхности звезд (скорость освобождения на поверхности Солнца (21), а также скорость движения галактик относительно реликтового фонового излучения $V_G$=600 км/с [33]). Солнце и его система двигаются относительно реликтового фонового излучения со скоростью около 400 км/с. И она не может превышать теоретический предел $V_3/\sqrt{2}$ =436.38 км/с.

По существу, скорость $V_3$ является предельной скоростью для любых конденсированных объектов. Звезды в этом случае являются "черными дырами" для конденсированного вещества. В отличие от "черных дыр" Лапласа–Эйнштейна в нашем случае их свободно покидают фотоны, но не может покинуть ни одно конденсированное тело, то есть, космический корабль априори не сможет взлететь с любой звезды: он просто испарится даже без внешнего нагрева. В этом случае конденсированное тело, например, в виде зародыша планеты, может покинуть звезду только при ее сжатии.

Движения звезд вокруг центра галактик также будут ограничены скоростями

$$V_4 = \frac{\alpha c}{8} = 273.46 \text{ км/с}. \qquad (55)$$

В нашей Галактике максимальная скорость движения звезд относительно центра находится на расстоянии 9,5 кпс от центра и равняется 273 км/с [23, с.1215].

Движение близлежащих звезд в галактиках относительно друг друга ограничено скоростью:

$$V_4 = \alpha^2 c = 15.964 \text{ км/с}. \qquad (56)$$

Например, скорость движения Солнца относительно ближайших к нам звезд равна 15.5 км/с [23] и не превышает значения (56).

Следующими за солнечной системой слоями пространств, которые подчиняются макроквантовым законам, являются квазары и ядра галактик. Однако достоверность экспериментальной информации о них недостаточна. Поэтому обратимся к более крупным масштабам.

По аналогии с солнечной системой найдем, как при выводе (8), протонную компоненту расслоения галактик относительно их центра при условии, что конец вектора электростатического поля протона движется со скоростью $\alpha^{-4}c$:

$$R_{Ge} = \frac{c}{\alpha^4 f_p} = \frac{h}{\alpha^{16} m_p c} = 2.043 \cdot 10^{19} \text{ м} = 0.6622 \text{кпс}, \qquad (57)$$

и электронную компоненту расслоения, как при выводе (11), при условии, что конец вектора электростатического поля электрона движется со скоростью $\alpha^{-4}c$:

$$R_{Ge} = \frac{c}{\alpha^4 f_e} = \frac{h}{\alpha^{15} m_e c} = 2.738 \cdot 10^{20} \text{ м} = 8.87 \text{кпс}. \qquad (58)$$

Такие расстояния соответствуют максимальным скоростям (55) движения звезд вокруг центра галактик. Однако в них, в отличие от солнечной системы, отсутствует четкое расслоение. По-видимому, как установил Финзи [34], это связано с тем, что на расстояниях более 1 кпс начинает нарушаться закон тяготения Ньютона. Это не позволяет галактикам группироваться во вращающиеся системы сверхгалактик наподобие обычных галактик. Следовательно, можно утверждать, что на расстояниях более 1 кпс, ньютоновский закон с постоянной $G$ должен переходить в макроквантовые законы, в которых допускается уменьшение или увеличение $G$. Прямым следствием совпадения экспериментальных данных и теоретических вычислений по (57), (58) является ограничение скорости гравитационных взаимодействий в метагалактике величиной $\alpha^{-4}c = 3.526 \cdot 10^8 c$. Эта скорость в 50 раз превышает оценку Лапласа для солнечной системы, который оценил ее величиной более $7 \cdot 10^6 c$.

Выразим постоянную Хаббла через электромагнитные константы:

$$H_0 = \frac{\alpha^{18} m_e c^2}{\hbar} = 82.489 \text{ км с}^{-1} \text{Мпс}^{-1}. \qquad (59)$$

В этом виде ее можно считать новой мировой константой. В разных источниках значение постоянной Хаббла оценивается в диапазоне от 50 до 100 км с$^{-1}$Мпс$^{-1}$ [6].

В связи с тем, что видимая Вселенная изотропна как по распределению галактик, так и по реликтовому и микроволновому излучению, Метагалактику можно считать сферической [15]. В этом случае значение постоянной Хаббла может являться квантовой частотой вращения Метагалактики как целого объекта, имеющего радиус

$$R_M = \frac{c}{H_0} = \frac{\hbar}{\alpha^{18} m_e c} = 1.1214 \cdot 10^{23} \text{ км}. \qquad (60)$$

Делая такие серьезные обобщения, определим характер нашего движения в пространстве (оно абсолютное или относительное). Так, в ОТО оно относительное, и, следовательно, находясь в космическом корабле, нельзя определить его скорость без внешних ориентиров.

После открытия изотропного реликтового излучения вопрос должен был бы решиться сам собой в пользу абсолютного движения, ибо появился некий эфир. И действительно, по доплеровскому смещению реликтового излучения были определены как скорость движения Солнца, так и скорость движения галактик относительно реликтового излучения, которые не превышают значения (54).

За счет полной изотропии в распределении галактик можно говорить об идентичности всех галактик, а о фоновом излучении как изотропном. Следовательно, абсолютная скорость движения галактик относительно фонового излучения не будет превышать значения (54).

Необходимо также отметить, что абсолютное движение космических тел в пространстве вызывает их деформацию [28], и они приобретают грушевидную форму по направлению их движения. Например, деформация Солнца и планет солнечной системы происходит не по направлению движения вокруг центра галактики, а в направлении апекса движения солнечной системы в целом [28, 35]. Важно, что деформация гравитационной массы Земли и ее магнитного поля происходит в одном и том же направлении [28], что экспериментально свидетельствует о непосредственной связи гравитационного и электромагнитного полей. Следовательно, любой наблюдатель-космонавт может только по деформации своего корабля определить его направление и скорость вплоть до скоростей, при которых корабль полностью испарится.

Если с новых позиций возвратиться к экспериментам Майкельсона–Морли по определению скорости движения Земли относительно эфира, которые лежат в основе ОТО, то интерферометр должен иметь базу, по крайней мере, не меньше размера солнечной системы или иметь скорость значительно выше скорости движения Земли, чтобы заметить эффект абсолютности движения.

Пожалуй, последнее серьезное противоречие предлагаемой теории связано с экспериментальным определением красного смещения галактик. Эксперимент показал, что, чем дальше объекты находятся друг от друга, тем больше красное доплеровское смещение в оптическом диапазоне [23]. На первом этапе красное смещение галактик хорошо вписывалось в теорию Большого взрыва, вытекающую из ОТО. Однако последующие эксперименты определили, что скорость разбегания дальних галактик превышает скорость света вплоть до 3$c$. Этот противоречащий ОТО факт постоянно игнорируется, что нарушает принципы методологии науки.

Этот парадокс разрешается следующим образом. Представим покраснение фотонов как результат их взаимодействия с межгалактическим гравитационным полем. Первые попытки решить этот парадокс таким образом были предприняты в работе [36]. Следуя ей, запишем красное смещение фотонов при движении их в межгалактическом пространстве за время *t* в следующем упрощенном виде:

$$\Delta W_g = \frac{\alpha^8 h\nu}{tH_0}, \qquad (61)$$

где ν – частота фотона.

Для расстояний, равных радиусу Метагалактики, предельное покраснение фотонов составит

$$\Delta W_g = 3.0082\ h\nu. \qquad (62)$$

Такому красному смещению соответствует красное смещение наиболее удаленных галактик. Однако оно для наиболее удаленных квазаров оказалось даже несколько больше $Z = 4.04$ [23, с.1224;33] По-видимому, это связано с тем, что фотоны, испущенные квазаром, претерпевают дополнительное красное смещение, вызванное его большой гравитационной массой.

Ряд других факторов, свидетельствующих о стационарности Вселенной, то есть об отсутствии разбегания галактик, а следовательно, и доплеровского красного смещения, хорошо представлен в [37].

### *Выводы*

1. С использованием "виртуального" измерительного средства методом фундаментальной квантовой выборки астрономических экспериментальных данных найдены законы, объединяющие электромагнитное, гравитационное и сильное взаимодействия.
2. Основываясь на геометрии пространства, удалось найти новые фундаментальные константы расслоения пространства, определяемые (2), (3).
3. Из этих констант вытекают новые безразмерные константы, например, важнейшая из них, ограничивающая естественным образом таблицу Менделеева максимальным зарядом ядра – зарядом ядра урана. Это делает бессмысленным поиск устойчивых трансурановых элементов.
4. На основании простой функциональной зависимости констант электромагнитного взаимодействия построена шкала для «виртуального» прибора, позволяющая измерять как электромагнитные, так и гравитационные взаимодействия (4).
5. Показано, что, используя безразмерную перенормировку вида $\alpha^{\pm n}$, можно привести планковские энергию, длину, массу, время к известным экспериментальным результатам, не нарушая общую идею Планка.
6. Найдены макроквантовые законы семьи Солнца и показано, что, в отличие от ньютоновской и эйнштейновской теорий, орбиты планет строго квантованы.
7. Найдены макроквантовые законы самого Солнца. Показано, что ряд необъяснимых ранее экспериментальных факторов достаточно просто находится с помощью модели полого Солнца. В частности, энергия Солнца имеет не термоядерное происхождение, а гравитационное. Это энергия анизотропного высокотемпературного гравитационного сжатия плазменной оболочки Солнца.
8. Найдены макроквантовые законы солнечной системы в целом. Показано, что долговременное (миллиарды лет) устойчивое движение планет определяется плазменными волнами в оболочке Солнца. Здесь потенциальная энергия сжатия оболочки переходит посредством гравитационных полей в кинетическую энергию движения планет по квантованным орбитам вокруг Солнца.
9. Выражена гравитационная константа через электромагнитные константы и массу протона.
10. Показано, что скорость распространения гравитационных взаимодействий конденсированное тело – электромагнитная волна полностью совпадает с эйнштейновской оценкой и равняется скорости света. Предельная скорость гравитационного взаимодействия конденсированного вещества между собой в нашей Галактике равняется $\alpha^{-4}c = 3.526 \cdot 10^8 c$.
11. Найдены макроквантовые законы для Земли. Показано, что ее внешнее ядро является солнечным веществом, потенциальная энергия сжатия которого переходит в кинетическую энергию вращения Земли, кинетическую энергию вращения Луны вокруг Земли, кинетическую энергию движения воды в океанах, тепловую энергию нагрева мантии. Климат на Земле, наряду с солнечной световой энергией, определяется движением волн во внешнем ядре Земли. Эти волны посредством индукционных взаимодействий передают энергию мантии, коре, океанам и атмосфере.
12. Найдены макроквантовые законы движения в Метагалактике. Определены предельные скорости движения звезд относительно друг друга и центра Галактики, скорости движения галактик относительно реликтового излучения.
13. Выражена постоянная Хаббла через электромагнитные константы, что позволяет использовать ее как новую мировую константу в виде частоты вращения метагалактики как целого объекта.
14. Показано, что максимально достижимая скорость движения конденсированных объектов в галактиках и в Метагалактике является абсолютной относительно реликтового излучения и в принципе не может превышать скорость $\alpha c$. В этом случае законы физики для конденсируемых веществ не инвариантны к преобразованию Лоренца.
15. Отмечено, что межпланетные путешествия землян и инопланетян требуют огромного времени, и поэтому такие путешествия теряют всякий смысл. По существу, мы обречены на физическое одиночество. Единственный путь общения с инопланетянами – это прямой обмен информацией со скоростью значительно выше скорости света.